# Magnetic susceptibility, magnetization, magnetic moment and characterization of Carancas meteorite

Domingo Rosales[*]
Erick Vidal[**]

**Resumen**

El 15 de Setiembre del 2007, en la comunidad de Carancas (Puno, Perú) un meteorito rocoso formo un cráter de tipo explosivo con un diámetro promedio de 13.5 m. varias muestras de fragmentos de meteorito fueron coleccionados. El análisis petrológico realizado indica que corresponde a un meteorito del tipo condrito ordinario H 4-5.

En el presente trabajo analizamos las propiedades magnéticas de un fragmento del meteorito con un magnetómetro de protones.

Con la finalidad de tener una completa caracterización del meteorito de Carancas y su cráter, a partir de diversas publicaciones, artículos y reportes, realizamos una compilación de las más importantes características y propiedades de este meteorito.

Palabras clave*: meteorito de Carancas, magnetización remanente, susceptibilidad magnética.*

**Abstract**

On September, 15$^{th}$, 2007, in the community of Carancas (Puno, Peru) a stony meteorite formed a crater explosive type with a mean diameter of 13.5 m. some samples meteorite fragments were collected. The petrologic analysis performed corresponds to a meteorite ordinary chondrite H 4-5.

In this paper we have analyzed the magnetic properties of a meteorite fragment with a proton magnetometer.

\* Observatorio geomagnético de Huancayo, Instituto Geofísico del Perú, domingo.igp@gmail.com
\*\* Observatorio geomagnético de Huancayo, Instituto Geofísico del Perú, erick.vidal@igp.gob.pe

Also in order to have a complete characterization of the Carancas meteorite and its crater, from several papers, articles and reports, we have made a compilation of the most important characteristics and properties of this meteorite.

**Introduction**

On September, 15$^{th}$, 2007, 11:40:14.4 Local Time, close to noon, (16:40:14.4 UT), the impact of a stony meteorite took place in the community of Carancas, Desaguadero town, Chucuito city, Region of Puno, Peru, on the south, near to lake Titicaca on the border with Bolivia, forming a crater explosive type with a mean diameter of 13.5 m. The geodesic coordinates of the center of the crater measured with GPS are: latitude 16°39' 52.2" South, longitude 69°02' 38.8" West, altitude 3,824 m. Meteorite fragments were sent to laboratories for petrologic analysis; results correspond to a meteorite type Ordinary Chondrite and of group H 4-5 by their iron content.

In this paper we have analyzed the magnetic properties (remanent magnetization, remanent magnetic moment, magnetic susceptibility and induced magnetic moment) of a meteorite fragment with a Overhauser proton magnetometer POS-1.

On the other hand, spectrographic, chemical and others analyses were performed. Several papers, articles and reports were published in different journals, each work showing different characteristics and properties such as the fireball and its trajectory, the meteorite, the crater and the ejecta. In order to have a complete characterization, we have made a compilation of all these works of the most important characteristics of the Carancas meteorite and its crater.

**Magnetic susceptibility, magnetization and magnetic moment measurements**

Magnetic susceptibility and magnetization of rocks and the permanent and induced moment of objects such as a meteorite can be measured using a proton magnetometer. The procedure involves rotating a sample about a point close to the magnetometer sensor on a line which is in the direction of the earth's total field $F$ and passes through the center of the sensor. Measurements of the maximum anomaly $T_{max}$ and minimum anomaly $T_{min}$ observed and the value of the field without the sample present $T_0$ is sufficient to allow calculation of magnetic susceptibility and induced and remanent magnetization. Next, measure the diameter of the sample which should be spherical as possible, measure the average diameter $D$, of the specimen and the distance $r$, between the center of the specimen when rotated and the center of the sensor. These five parameters, $T_0$, $T_{max}$, $T_{min}$, $D$, and $r$, are all that is needed in the following formulae to calculate both magnetic susceptibility and magnetization or the induced and permanent magnetic moments of a small object (Breiner, 1973).

**Mathematical formulations**

For the remanent magnetization:
The remanent anomaly $T_r$ is given by:

$$T_r = \frac{T_{max} - T_{min}}{2} = \frac{2M_r}{r^3} = \frac{2I_r \frac{4}{3}\pi\left(\frac{D}{2}\right)^3}{r^3} \quad (1)$$

Therefore

$$I_r = \frac{3}{2\pi}\left(\frac{r}{D}\right)^3 (T_{max} - T_{min}) \quad (2)$$

and

$$M_r = \frac{r^3}{4}(T_{max} - T_{min}) \quad (3)$$

where $I_r$ is the remanent magnetization per unit volume expressed in $nT$ $(1 \cdot nT = 10^{-5} G)$[1] and $M_r$ is the remanent magnetic moment expressed in "electromagnetic unit ($emu$)" $(1 \cdot emu = 1 \cdot G \cdot cm^3)$.

For the induced magnetization:
The induced anomaly $T_i$ is given by:

$$T_i = \frac{T_{max} + T_{min}}{2} - T_0 = \frac{2M_i}{r^3} = \frac{2I_i \frac{4}{3}\pi\left(\frac{D}{2}\right)^3}{r^3} \quad (4)$$

and

---

[1] $G$ : Gauss

$$I_i = kF = \frac{3}{2\pi}\left(\frac{r}{D}\right)^3 (T_{max} + T_{min} - 2T_0) \tag{5}$$

Hence

$$k = \frac{3}{2\pi F}\left(\frac{r}{D}\right)^3 (T_{max} + T_{min} - 2T_0) \tag{6}$$

and

$$M_i = \frac{r^3}{4}(T_{max} + T_{min} - 2T_0) \tag{7}$$

where $I_i$ is the magnetic susceptibility per unit volume expressed in $nT$, and $M_i$ is the induced magnetic moment expressed in $emu$. $k$ is the magnetic susceptibility, is a dimensionless proportionally constant that indicates the degree of magnetization of a material in response to an applied magnetic field. The total magnetization per unit volume is $I_t = I_r + I_i$.

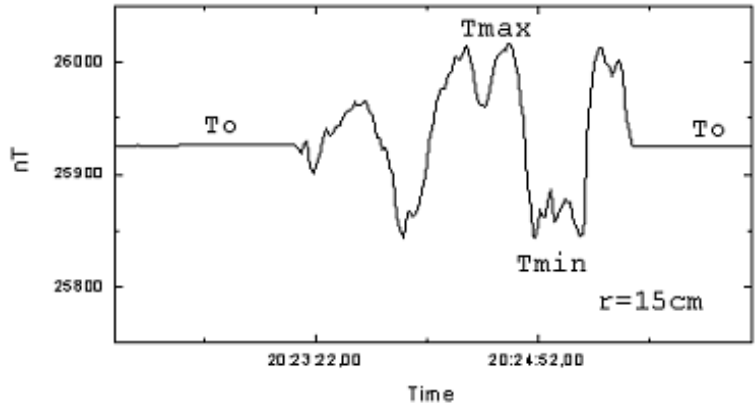

**Figure 1.** Readigns during rotation of meterorite fragment to 15cm. away from the proton sensor.

The meteorite fragment analysed have 28.777 gr., an average diameter $D = 2.460$ cm, with density 3.629 gr/cm³. The readings obtained during rotation of specimen to 15cm away from the proton sensor are shown in Figure 1.

Eight sets of measurements were performed at 10, 15, 20, 25, 30 40, 50 and 75 cm. away from the proton sensor, the set at 10 cm was removed due to the signal of proton magnetometer was degraded.

The determination of magnetic properties of Carancas meterorite is given in the table 1. For $I_i$ and $M_i$ has not been possible to determine with greater precision so these values are only referentially.

**Table 1**
**Magnetic susceptibility and induced and remanent magnetization**

| r(cm) | $T_r$ (nT) | $T_i$ (nT) | $I_r$ (G) | $M_r$ (emu) | $I_i$ (G) | $M_i$ (emu) |
|---|---|---|---|---|---|---|
| 15 | 93.132 | 9.094 | 0.2016 | 1.5720 | 0.0197 | 0.1530 |
| 20 | 39.710 | 6.393 | 0.2038 | 1.5890 | 0.0328 | 0.2560 |
| 25 | 20.245 | 0.514 | 0.2029 | 1.5820 | 0.0052 | 0.0400 |
| 30 | 11.569 | 0.863 | 0.2004 | 1.5620 | 0.0149 | 0.1170 |
| 40 | 4.871 | 0.658 | 0.2000 | 1.5590 | 0.0157 | 0.1220 |
| 50 | 2.671 | 0.658 | 0.2142 | 1.6690 | 0.0528 | 0.411 |
| 75 | 0.795 | 0.402 | 0.2151 | 1.6770 | - | - |
| **Mean** | | | **0.2054±0.0064** | **1.6014±0.0500** | **0.0235±0.0169** | **0.1832±0.1317** |

**Characterization of Carancas meteorite and its crater**

Some samples meteorite fragments were collected. Magnetic, spectrographic, chemical and others analysis were performed. Several papers and articles about Carancas meteorite were published, each work showing different characteristics and properties. The compilation of all these works with the most important characterizations of Carancas meteorite and its crater are:

**The place impact:**

| Latitude | 16°39' 52.2" South | (Rosales et al., 2008) |
|---|---|---|
| Longitude | 69°02' 38.8" West | |
| Altitude | 3,824 m. | |
| Impact date | September 15$^{th}$, 2007 | |
| Impact time UT | 16:40:14.4 universal time | (Tancredi et al., 2009) |
| Impact time LT | 11:40:14.4 local time | |
| Community | Carancas | (Rosales et al., 2008) |
| City | Chucuito | |
| Region | Puno | |
| Country | Peru | |

**Meteoroid before entering the atmosphere:**

| Initial velocity | Between 12 – 17 km/s | (Tancredi et al., 2009) |
|---|---|---|
| Initial mass | Between 7 – 12 ton | |
| Initial diameter | Between 1.6 – 2.0 m | |
| Initial kinetic energy | From 0.12 – 0.41 kT TNT | |
| Original orbit meteoroid | Compatible with NEAs[2] | |

**The meteoroid before the impact:**

| Diameter | Between 0.6 - 1.1 m | (Tancredi et al., 2009) |
|---|---|---|
| Mass | Between 0.3 – 3.0 tons | |

**The meteoroid during the impact:**

| Impact velocity | Between ~3 - 6 km/s | (Tancredi et al., 2009) |
|---|---|---|
| Impact angle | Between 45° - 60° | |
| Trajectory azimuth | Between 80° - 110° | |
| Impact energy | Between ~1 – 3 tons TNT | |
| Local seismic magnitude | ML = 1.45 | (Le Pichon et al., 2008) |
| Seismic energy generated by impact | Roughly $9.4 \times 10^6$ J Equivalent to 2.3 kg TNT | (Tancredi et al., 2009) |
| Seismic efficiency | $10^{-3}$ | |

**The meteorite:**

| Type | Ordinary Chondrite H4-5 | (Varela and Branztatter, 2007) |
|---|---|---|
| Density | 3.629 gr/cm$^3$ | (Rosales et al., 2008) |
| Remanent magnetization per volume unit | $I_r = 0.2054$ Gauss | |
| Ablation coefficient | 0.014 s$^2$/km$^2$ | (Kenkmann et al., 2008) |

**The crater:**

| Diameter approxi- | 13.5 m | (Rosales et al., 2008) |
|---|---|---|

---

[2] NEAs: Near-Earth Asteroids

| mately | | |
|---|---|---|
| Deep of crater | Between 2.4 – 5.0 m | (Tancredi et al., 2008) |
| Type crater | Explosive | (Pereira, 2007) |
| Shape of the crater | Nearly circular | |
| Diameter water pond in the crater | Between 7.4 – 7.8 m | (Miura, 2008) |

**The ejecta:**

| Density of ejecta | 1.700 gr/cm$^3$ | (Tancredi et al., 2009) |
|---|---|---|
| Ejection max. distance and direction | 348 m SW | (Rosales et al, 2008) |

**The meteorite mineralogical composition:**

| Pyroxene 1 | 40 % | |
|---|---|---|
| Olivine | 20 % | |
| Feldspar | 10 % | |
| Pyroxene 2 | 10 % | (Macedo et al., 2007) |
| Kamacite | 15 % | |
| Troilite | 5% | |
| Cromite | traces | |
| Native Cu | traces | |

**The meteorite chemical compound:**

| Compound | Ol (%) | Px (%) | |
|---|---|---|---|
| SiO$_2$ | 38.30 – 39.70 | 55.40 – 56.50 | |
| Al$_2$O$_3$ | 0.00 | 0.15 – 1.55 | |
| TiO$_2$ | 0.00 – 0.06 | 0.06 – 0.29 | |
| Cr$_2$O$_3$ | 0.00 – 0.03 | 0.14 – 0.76 | (Varela and Brandztatter, 2007) |
| MnO | 0.42 – 0.48 | 0.47 – 0.48 | |
| FeO | 17.1 – 17.2 | 16.60 | |
| CaO | 0.00 | 0.54 – 1.08 | |
| MgO | 43.40 – 43.50 | 29.10 – 31.40 | |

**The meteorite chemical elements:**

| | | |
|---|---|---|
| Si | 18 % | |
| Mg | 14 % | |
| Fe | 14 % | |
| Al | 1.7 % | |
| S | 1.6 % | |
| Ca | 1.5 % | (Nuñez del Prado et al., 2008) |
| Na | 1.9% | |
| Cr | traces | |
| P | traces | |
| K | traces | |
| Cu | traces | |
| Re | traces | |
| Rh | traces | |

**The oxygen isotope analysis:**

| | | |
|---|---|---|
| $ϭ^{17}O$ | Between 3.02 – 2.94 % | |
| $ϭ^{18}O$ | Between 4.52 – 4.32 % | (Nuñez del Prado et al., 2008) |
| $\Delta^{17}O$ | Between 0.67 – 0.68 % | |

**The X-ray diffractometry analysis:**

| | | |
|---|---|---|
| Pyroxene | paramagnetic | |
| Olivine | paramagnetic | (Cerrón and Bravo, 2008) |
| Troilite(FeS) | magnetic | |
| Kamacite (FeNi) | magnetic | |

**Conclusion:**

The Carancas meteorite is a ordinary chondrite H 4-5, which produced a crater on Earth's surface, an event like this should have not occurred, normally such meteoroids ablating on the Earth's atmosphere.

Carancas crater was product of a hypervelocity impact. In spite of significant ablation, the meteoroid did not catastrophically disrupt and/or disperse during its atmospheric entry (Tancredi et al., 2009).

Carancas meteorite has a remanent magnetization of $I_r = 0.2054$ Gauss and remanent magnetic moment $M_r = 1.6014$ emu. Induced magnetization could not be determined with greater precision so this value determinates ($I_i \approx 0.0235$ Gauss, and $M_i \approx 0.1832$ emu) are only referentially.

The characterization of the Carancas meteorite and its crater allows us to have all the necessary information about this event.